\documentclass[a4paper,12pt]{article}
\usepackage{epsf,amsmath}
\usepackage{graphicx}
\usepackage{amssymb}
\usepackage{array,tabularx}
\usepackage{indentfirst}
\usepackage[utf8]{inputenc}
\usepackage{cite}
\begin{document}

\title{Stability in higher-derivative matter fields theories.}

\author{Petr V. Tretyakov}
\date{}
\maketitle
{\it \begin{center}Bogoliubov Laboratory of Theoretical Physics, Joint Institute for Nuclear Research\end{center}
\begin{center}Joliot-Curie 6, 141980 Dubna, Moscow region, Russia\end{center}
\begin{center}and\end{center}
\begin{center}Department of General Relativity and Gravitation, Institute of Physics,\end{center}
\begin{center}Kazan Federal University, Kremlevskaya street 18, 420008 Kazan, Russia\end{center}
\begin{center}tpv@theor.jinr.ru\end{center}}

\begin{abstract}
We discuss possible instabilities in higher-derivative matter fields theories. These theories has two free parameters $\beta_1$ and $\beta_4$. By using dynamical system approach we explicitly demonstrate that for stability of Minkowski space in expanding Universe it is need condition $\beta_4<0$. By using quantum field theory approach we also find additional restriction for parameters $\beta_1>-\frac{1}{3}\beta_4$ which is need to avoid tachyon-like instability.


\end{abstract}

\section{Introduction}
The unsolved problems of General Relativity such as dark energy \cite{R,P} and dark matter \cite{Planck} force us to investigate different alternatives. One of the most popular alternative is well known $f(R)$-gravity \cite{BCNO,NO} in the different forms \cite{CM,A,NOO}. There are also a number of another possibilities to modify gravity, such as Palatini $f(R)$-gravity \cite{HKLO}, teleparallel gravity \cite{AP}, Horndenski theory \cite{KYY}, theories with non-minimal kinetic coupling \cite{Sushkov} and so on \cite{CFPS}.

A new type of modified gravity was recently proposed \cite{PSV}. The higher derivative matter fields are implied in such kind of theory, which can be interpreted as non-dynamical auxiliary fields.

\begin{equation}
G_{\mu\nu}+\Lambda g_{\mu\nu}= T_{\mu\nu}+ S_{\mu\nu}(\mathbf{g},\mathbf{T}),
\label{1.1}
\end{equation}

The most general form of the term $S_{\mu\nu}$ if we take into account terms up to fourth order in derivatives is

\begin{equation}
\begin{array}{r}
S_{\mu\nu}=\alpha_1 g_{\mu\nu}T + \alpha_2 g_{\mu\nu} T^2 + \alpha_3 T T_{\mu\nu}+ \alpha_4 g_{\mu\nu}T^{\alpha\beta}T_{\alpha\beta}+\alpha_5 T^{\alpha}_{\,\,\,\,\,\mu}T_{\alpha\nu}\\
\\+ \beta_1\nabla_{\mu}\nabla_{\nu}T + \beta_2g_{\mu\nu}\Box T+ \beta_3\Box T_{\mu\nu}+2\beta_4\nabla^{\alpha}\nabla_{(\mu}T_{\nu)\alpha},
\end{array}
\label{1.2}
\end{equation}
where since we impose divergence free condition $S^{\mu\nu}_{;\mu}=0$\footnote{Note that this imply $T^{\mu\nu}_{;\mu}=0$ as well.} there are a number of relations between the coefficients. So finally we have only two independent parameters $\beta_1$ and $\beta_4$ and equation (\ref{1.1}) takes the form \cite{PSV}

\begin{equation}
\begin{array}{r}
G_{\mu\nu}= T_{\mu\nu} -\Lambda g_{\mu\nu} -\beta_1\Lambda g_{\mu\nu}T + \frac{1}{4}(1-2\beta_1\Lambda)(\beta_1-\beta_4) g_{\mu\nu} T^2 + [\beta_4(1-2\beta_1\Lambda)-\beta_1] T T_{\mu\nu}\\
\\+ \frac{1}{2}\beta_4 g_{\mu\nu}T^{\alpha\beta}T_{\alpha\beta}-2\beta_4 T^{\alpha}_{\,\,\,\,\,\mu}T_{\alpha\nu}
+ \beta_1\nabla_{\mu}\nabla_{\nu}T - \beta_1g_{\mu\nu}\Box T - \beta_4\Box T_{\mu\nu}+2\beta_4\nabla^{\alpha}\nabla_{(\mu}T_{\nu)\alpha}.
\end{array}
\label{1.3}
\end{equation}

Some of the terms from equation (\ref{1.3}) also appears in $f(R,T)$-theory \cite{HLNO}.
Note that for special choice of parameters some of well known theories are contained in representation (\ref{1.3}) as a limit. For instance case $\beta_1=0$, $\beta_4=-\kappa/2$ corresponds to EiBI gravity in the small coupling limit or $\beta_4=0$ correspond to generic Palatini $f(R)$ gravity \cite{PSV}. In this sense theory (\ref{1.1}) also may be interpreted as some kind of phenomenological theory of modified gravity. Cosmology in such kind of theories was investigated in \cite{HLS,BBC}. Generalizations for brane theories was studied in \cite{GLY,BLM,BMMM}. Some interesting remarks about this theory may be found also in \cite{BBCG}.

For standard perfect fluid $T_{\mu\nu}=(p+\rho)U_{\mu}U_{\nu}+ p g_{\mu\nu}$ and FLRW metric $ds^2=-dt^2+a^2(t) \delta_{\mu\nu}dx^{\mu}dx^{\nu}$ Friedman-like equation takes the form \cite{HLS}

\begin{equation}
\begin{array}{r}
3H^2= \rho+\Lambda +3H[\beta_1(\dot\rho-3\dot p)-\beta_4(\dot\rho+2\dot p)] -\beta_4\ddot\rho +\frac{1}{4}\left[3\beta_1(1+2\beta_4\Lambda)(\rho^2-3p^2)\right. \\
\\ \left.+ 3\beta_4(\rho^2+p^2) -12\beta_1(\beta_1+\beta_4)\Lambda\rho p - 6(\beta_1-\beta_4)\rho p -4\beta_1\Lambda(\rho-3p) +2\beta_1^2\Lambda (\rho^2+9p^2) \right],
\end{array}
\label{1.4}
\end{equation}
and also we have usual energy conservation law $T^{\mu\nu}_{;\mu}=0$ which reads now
\begin{equation}
\dot\rho+3H(p+\rho)=0,
\label{1.5}
\end{equation}
where $H=\frac{\dot a}{a}$ is usual Hubble parameter.

\section{Stability conditions}

It is well known that incorporating of higher derivatives terms can affect on stability of different solutions including simplest cosmologically important. One of the most important is Minkowski solution. Let us study stability of Minkowski solution in such kind of gravity. To discuss stability problem we need to transform equations to dynamical system form. First of all we will discuss the simplest (and the most convenient for cosmological applications) equation of state $p=w\rho$. Further we can express Hubble parameter from equation (\ref{1.5}) and insert it to (\ref{1.4}). By this way we obtain second order differential equation for function $\rho$, which is also depend on parameters $\beta_1$, $\beta_4$, $\Lambda$ and $w$:
\begin{equation}
\beta_4\ddot\rho=\rho +\Lambda +\frac{1}{4}\rho^2f_2 - \frac{f_1}{1+w}\frac{\dot\rho^2}{\rho} - \frac{1}{3(1+w)^2}\frac{\dot\rho^2}{\rho^2},
\label{2.1}
\end{equation}
with
\begin{equation}
f_1=\beta_1(1-3w)-\beta_4(1+2w),
\label{2.2}
\end{equation}
and
\begin{equation}
\begin{array}{r}
f_2=3\beta_1(1-3w^2)(1+2\beta_4\Lambda)+3\beta_4(1+w^2) -12\beta_1(\beta_1+\beta_4)\Lambda w -6w(\beta_1-\beta_4)\\
\\ -4\beta_1(1-3w)\Lambda +2\beta_1\Lambda(1+9w^2).
\end{array}
\label{2.3}
\end{equation}
We can see that $f_1$ and $f_2$ are non dynamical functions and only depend on parameters of the theory.

For further investigation let us rewrite equation (\ref{2.1}) as dynamical system
\begin{equation}
\left\{
\begin{array}{r}
\dot\rho=\pi,\\
\\ \dot\pi = F,
\end{array}
\right.
\label{2.4}
\end{equation}
with function $F$
\begin{equation}
F(\rho,\pi)\equiv \frac{1}{\beta_4}\left (\rho+\Lambda+\frac{1}{4}\rho^2 f_2 - \frac{f_1}{1+w}\frac{\pi^2}{\rho} - \frac{1}{3(1+w)^2}\frac{\pi^2}{\rho^2} \right ).
\label{2.5}
\end{equation}
It is clear that stationary points of system (\ref{2.4}) is governed by the next simple equation
\begin{equation}
\frac{1}{4}f_2\rho^2+\rho+\Lambda=0.
\label{2.6}
\end{equation}
Now let us discuss the physical meaning of its solutions. First of all note that we are interesting in solutions with true vacuum $\Lambda=0$, but we need to keep parameter $\Lambda$ to investigate perturbations because it coupled with another parameters of our theory $\beta_1$ and $\beta_4$. So we can to vanish it only at the end of our investigation. So we may interpret $\Lambda$ as parameter which allow us to avoid degeneration of solutions and we can put it $\Lambda\rightarrow 0$ at the end. Similar approach was successfully practiced to stability investigation of Minkowski solution in different modified gravity theories in our previous papers \cite{T1,T2}. First solution of equation (\ref{2.6}) reads (\ref{2.6}) reads
\begin{equation}
\rho=\frac{-2-2\sqrt{1-f_2\Lambda}}{f_2}.
\label{2.7}
\end{equation}
We can see that in the case $\Lambda\rightarrow 0$ this solution reads $\rho=-\frac{4}{f_2}\neq 0$, moreover (\ref{1.5}) tell us that must be $H=0$. So this solution corresponds to the static Einstein universe and it's not interesting for our further investigations, but if we want to keep it, we must put $f_2<0$ to satisfy week energy condition ({\it wek})\footnote{Note that in our case $p=w\rho$, $w\neq -1$ and {\it wec} directly follows from null energy condition ({\it nec}).} and some additional restrictions for parameters will follows from this inequality\footnote{It reads $\beta_1<\frac{\beta_4(w+1)}{3w-1}$ or for the dust($w=0$) case $\beta_1<-\beta_4$.}. Second solution reads
\begin{equation}
\rho=\frac{-2+2\sqrt{1-f_2\Lambda}}{f_2},
\label{2.8}
\end{equation}
and it have a limit $\rho \rightarrow -\Lambda$ for $\Lambda\rightarrow 0$, so we need to put $\Lambda \rightarrow -0$ to satisfy {\it wec}. Note that this solution exist for any value of $H$ including the case $H=0$. In the last case this is Minkowski solution and its stability is very important for us. Let us study Minkowski stability conditions. It is well known that in the first order stability governed by the equation

\begin{equation}
\left |
\begin{array}{l}
-\mu\,\,\,\,\,\,\,\,\,\,\,\,\,\,\,\,\,\,\,\,\,\,1\\
\\ (F_{\rho})_0\,\,\,\,\,\,(F_{\pi})_0-\mu
\end{array}
\label{2.9}
\right |=0,
\end{equation}
or\footnote{Here we imply $F_{\rho}\equiv\frac{\partial F}{\partial\rho}$ and so on.}
\begin{equation}
\mu^2-(F_{\pi})_0\mu - (F_{\rho})_0=0,
\label{2.10}
\end{equation}
and it easy to see that stability conditions takes the form\footnote{Remind that $Re(\mu_{1,2})<0$ needs for stability.}
\begin{equation}
\left \{
\begin{array}{l}
(F_{\pi})_0<0,\\
\\ (F_{\rho})_0<0.
\end{array}
\label{2.11}
\right.
\end{equation}
Let us calculate values of these functions explicitly. For function $F_{\pi}$ we have
\begin{equation}
F_{\pi}=\frac{-2\dot\rho}{(1+w)\rho}\frac{1}{\beta_4}\left[ f_1 +\frac{1}{3(1+w)\rho} \right ].
\label{2.12}
\end{equation}
First of all note that at the interesting point this function takes singular value, but we need only its sign. At the second we can put $1+w>0$ without loss of generality, because case $w=-1$ corresponds to cosmological term case, which is already incorporate in our equations. Now we can see from (\ref{1.5}) that $\frac{-2\dot\rho}{(1+w)\rho}=6H>0$ in expanding universe. Value of function $f_1$ is always finite, see (\ref{2.2}), whereas value of expression $\frac{1}{3(1+w)\rho}$ is infinite at the studying point ($\rho=0$) and positive, if we put quite natural conditions $1+w>0$ and $\rho>0$. So we find that total sign of expression (\ref{2.12}) at $(0,0)$ point is governed by the sign of parameter $\beta_4$ and for Minkowski stability must be $\beta_4<0$.

Now let us discuss second condition from (\ref{2.11}). We have
\begin{equation}
F_{\rho}=\frac{1}{\beta_4} + \frac{1}{2}\rho f_2 +\frac{f_1}{\beta_4(1+w)}\frac{\dot\rho^2}{\rho^2} + \frac{2}{3\beta_4(1+w)^2}\frac{\dot\rho^2}{\rho^3}.
\label{2.13}
\end{equation}
First of all note that near interesting point $(0,0)$ second term may be neglected because $f_2$ have always finite value (\ref{2.3}) and $\rho\rightarrow 0$, whereas first term is equal to non-zero constant. According to (\ref{1.5}) $\frac{\dot\rho^2}{\rho^2}=9H^2(1+w)^2\rightarrow 0$ for Minkowski solution, so the third term may be also neglected. And for the forth term we have $\frac{\dot\rho^2}{\rho^3(1+w)^2}=\frac{9H^2}{\rho}=3$ near the point $\rho=0$, see (\ref{1.4}). So finally we have
\begin{equation}
(F_{\rho})_0=\frac{3}{\beta_4},
\label{2.14}
\end{equation}
and we can see that for stability it is need $\beta_4<0$ as well.

Now let us discuss possible instabilities from another point of view. Trace of equation (\ref{1.3}) reads
\begin{equation}
(\beta_4+3\beta_1)\Box T +2\Lambda\left ( 2 +2\beta_1 T +\beta_1^2T^2 \right) -T-R =0.
\label{2.15}
\end{equation}
Let us study a small perturbations $\delta T$ under the some solution of this equation $T_0$ and $R_0$. So we put $T=T_0+\delta T$ and equation for $\delta T$ takes the form
\begin{equation}
(\beta_4+3\beta_1)\Box \delta T +4\Lambda\left ( \beta_1 +2\beta_1^2T_0 \right)\delta T - \delta T =0,
\label{2.16}
\end{equation}
we try to find possible solutions of (\ref{2.16}) as standard decomposition on functions
\begin{equation}
u_k\sim e^{i\mathbf{k}\mathbf{x}-i\omega t},
\label{2.17}
\end{equation}
where $\omega\equiv (k^2+\mu^2)^{1/2}$, $k\equiv\mid \mathbf{k}\mid$ and $\mu$ is the mass of effective scalar field (scalaron). After substituting this representation into (\ref{2.16}) we obtain next equation for $\mu^2$:
\begin{equation}
\mu^2=\frac{1-4\Lambda\left (\beta_1+2\beta_1^2T_0\right )}{\beta_4+3\beta_1},
\label{2.18}
\end{equation}
and if we turn back to the Minkowski limit $\Lambda\rightarrow 0$, this relation give us additional restriction to avoid tachyon-like instability ($\mu^2<0$):
\begin{equation}
\beta_1>-\frac{1}{3}\beta_4.
\label{2.19}
\end{equation}

\section{Conclusion}

In this paper we discuss instabilities in higher-derivative matter fields theories. We found condition for Minkowski stability and another one to avoid tachyon-like instability. Of course, these are only necessary but not enough conditions for stability of the theory. For instance we study Minkowski stability only with respect to the simplest class of isotropic perturbations and taking into account more of complicate perturbations may provide us some additional restrictions for parameters. Nevertheless the found conditions must be satisfied and it may be very helpful for further construction of the theory.

\section{Acknowledgments}

Author is grateful to A.A. Starobinsky for discussion and some useful remarks.
This work was supported by the Russian Science Foundation (RSF) grant 16-12-10401.

\end{document}